\documentclass[sigconf]{acmart}

\acmConference[ICSE 2024]{46th International Conference on Software Engineering}{April 2024}{Lisbon, Portugal}

\AtBeginDocument{%
  }

\copyrightyear{2024}
\acmYear{2024}
\setcopyright{acmlicensed}\acmConference[LLM4Code '24]{2024 International Workshop on Large Language Models for Code}{April 20, 2024}{Lisbon, Portugal}
\acmBooktitle{2024 International Workshop on Large Language Models for Code (LLM4Code '24), April 20, 2024, Lisbon, Portugal}
\acmDOI{10.1145/3643795.3648391}
\acmISBN{979-8-4007-0579-3/24/04}

\usepackage{color}

\definecolor{pblue}{rgb}{0.13,0.13,1}
\definecolor{pgreen}{rgb}{0,0.5,0}
\definecolor{pred}{rgb}{0.9,0,0}
\definecolor{pgrey}{rgb}{0.46,0.45,0.48}
\usepackage{listings}
\usepackage{multirow}

\begin{document}

\title{Toward a New Era of Rapid Development:
\\
Assessing GPT-4-Vision's Capabilities in UML-Based Code Generation
}

\author{Gábor Antal}
\email{antal@inf.u-szeged.hu}
\orcid{0000-0002-3002-8624}
\affiliation{%
	\institution{University of Szeged}
	\city{Szeged}
	\country{Hungary}
	\postcode{6720}
}

\author{Richárd Vozár}
\email{vozar.svab@gmail.com}
\affiliation{%
	\institution{University of Szeged}
	\city{Szeged}
	\country{Hungary}
	\postcode{6720}
}

\author{Rudolf Ferenc}
\email{ferenc@inf.u-szeged.hu}
\orcid{0000-0001-8897-7403}
\affiliation{%
  \institution{University of Szeged}
  \city{Szeged}
  \country{Hungary}
  \postcode{6720}
}

\renewcommand{\shortauthors}{Antal et al.}

\begin{abstract}
The emergence of advanced neural networks has opened up new ways in automated code generation from conceptual models, promising to enhance software development processes.
This paper presents a preliminary evaluation of GPT-4-Vision, a state-of-the-art deep learning model, and its capabilities in transforming Unified Modeling Language (UML) class diagrams into fully operating Java class files.
In our study, we used exported images of 18 class diagrams comprising 10 single-class and 8 multi-class diagrams.
We used 3 different prompts for each input, and we manually evaluated the results.
We created a scoring system in which we scored the occurrence of elements found in the diagram within the source code.
On average, the model was able to generate source code for 88\% of the elements shown in the diagrams.
Our results indicate that GPT-4-Vision exhibits proficiency in handling single-class UML diagrams, successfully transforming them into syntactically correct class files.
However, for multi-class UML diagrams, the model's performance is weaker compared to single-class diagrams.
In summary, further investigations are necessary to exploit the model's potential completely.


\end{abstract}




\maketitle

\section{Introduction}
\label{sec:intro}

In software engineering, the design phase is crucial in defining the system architecture and serves as a preliminary blueprint for development.
Unified Modeling Language (UML)~\cite{jacobson2021unified} has emerged as the de facto standard for modeling and visualizing the design of software systems.
UML provides various diagrams to express different aspects of software design, including structural, behavioral, and interaction diagrams.
These diagrams serve as a communication medium among various stakeholders, such as developers, analysts, and managers, and provide a basis for the implementation phase in the software development lifecycle. Moreover, when UML class diagrams are accurately and carefully designed, they can significantly reduce ambiguities, decreasing the likelihood of errors in the coding process.

The class diagram is particularly significant among the various types of UML diagrams.
It offers a static view of the system, detailing classes, their attributes, operations (or methods), and the relationships among them, all of which are essential for understanding object-oriented software design.
However, the process of manual transition from design to code is a time-consuming and resource-intensive task.
It often requires expertise in a particular programming language, and still, we are humans; we can make mistakes occasionally.

Automatic code generation from UML class diagrams could be useful in many cases.
For example, it can help in rapid application development by automatically creating classes the developer would not want to create.
It can also be useful if we want to check ourselves (while learning a programming language) what a given architecture would look like in a particular language. 
It may also be the case that only diagrams are available for part of the system and no source code (e.g., for legacy systems).
Developers might want to restore the non-existing classes with as little effort as possible.
Transforming UML automatically often requires the project files and not only the exported images.
Mainly, the code generation functions in UML editors generate source code with very limited functionality.

To address this, the concept of automatic code generation from UML seeks to revolutionize how developers work by automating the tedious and intricate task of writing boilerplate code.
Leveraging state-of-the-art natural language processing and computer vision technologies, such as GPT-4-Vision, promises to significantly reduce development time, minimize human error, and ease the transition from design to implementation. 
%
%
Of course, it would be highly beneficial if the generated source code would be a more or less functioning program that we could make fully operational with minor modifications.

To explore this possibility, the study examines OpenAI's latest GPT (Generative Pre-trained Transformer)\cite{attention} model enhanced with visual capabilities, the GPT-4-Vision model, which can process images.
As our early results confirmed, GPT-4-Vision is capable of interpreting UML diagrams.
To our knowledge, this is the first study that utilizes GPT-4-Vision for UML-based source code generation.
In our study, we examined UML class diagrams of various complexities, and with various prompts, we explored GPT-4-Vision's capabilities in source code generation, seeking answer to the following research question:

\begin{itemize}
  \item \textbf{RQ}: How effectively can GPT-4-Vision generate source code from UML diagrams of varying complexity?
\end{itemize}

The rest of the paper is organized as follows.
In Section~\ref{sec:related}, we list the related literature.
Section~\ref{sec:methodology} presents the used methodology.
We present and explain our preliminary results in Section~\ref{sec:results}.
We briefly outline our future plans in Section~\ref{sec:future}, and conclude the paper in Section~\ref{sec:conclusion}.

\section{Related Work}
\label{sec:related}

Creating source code from models has long been a standing research problem.
Many researchers have also attempted to recognize models using conventional methods.
One of the first works in this field was to recognize hand-drawn UML diagrams.
Hammond et al.~\cite{hammond2006tahuti} presented Tahuti, a sketch recognition environment for UML class diagrams.
Their system is based on a multi-layer recognition framework that recognizes objects by their geometrical properties.
Lank et al.~\cite{lank2000interactive} created a similar system to recognize hand-drawn UML diagrams.
Their system can recognize hand-drawn UML diagrams from various input devices through a sophisticated segmentation algorithm in a user-correctable interface.
Maggiori et al.~\cite{maggiori2014towards} presented the IMEAV framework to reverse-engineer module views from a variety of UML and non-UML diagrams.
Chen et al.~\cite{chen2022automatically} proposed the ReSECDI approach to automatically recognize the semantic elements from class diagram images using image processing technologies.
They evaluated their approach on 80 diagrams.
Karasneh et al.~\cite{karasneh2013extracting, karasneh2013img2uml} presented the Img2UML tool, which uses image processing as a basis.
Koziolek et al.~\cite{koziolek2023llm} proposed an LLM-based code generation method (which generates \textit{IEC 61131-3 Structure Text control logic source code}) from \textit{Piping-and-Instrumentation Diagrams} (P\&IDs) using image recognition.
Despite their field of study is not related to software engineering, the topic is quite similar to ours: they used ChatGPT to generate the source codes from the models and fed the generated to to OpenPLC to validate it manually.
They concluded that the proposed method is feasible, however, not perfect.
Han et al.~\cite{han2023comparative} investigated the capabilities of GPT-4-Vision in generating clinical diagnoses, where they compared the results achieved by textual and visual inputs.

\section{Methodology}
\label{sec:methodology}

At the beginning of our research, we collected hundreds of UML class diagrams from a variety of sources, e.g., students' assignments from university courses and numerous tutorials.
We categorized these diagrams according to their complexity.
In this study, we chose diagrams from only two categories:

\begin{itemize}
  \item \textbf{Single-class}: class diagrams containing only one class.
  \item \textbf{Multi-class}: class diagrams comprising multiple classes, featuring a mixture of abstract and concrete classes, interfaces, and inheritance relationships.
\end{itemize}

From these categories, we selected 10 examples of single-class diagrams and 8 examples of multi-class diagrams.
Subsequently, it was necessary to create appropriate input for the GPT-4-Vision model (i.e., \textit{prompt}); we used the UML class diagram and a textual prompt in each case.
Since this study aimed not to optimize textual prompts but to assess the code-generating capabilities of GPT-4-Vision, we conducted our experiment using relatively simple prompts.
However, we defined 3 levels of granularity for our prompts:
\begin{enumerate}
  \item We provided the simplest prompt and the corresponding UML class diagram. The prompt did not include any information about the project, classes, or functional details. In this case, we used the following prompt: ``\textit{Create Java source code from the image!}''
  \item The prompt concisely summarized the program represented by the given diagram. An example prompt is as follows: ``\textit{This is a really simple application where we have classes for books, reading, and bookshelf, and also got an exception for reading too fast. Create Java source code from the image! Try to implement the logic of these methods.}''
  \item The prompt detailed the program's purpose shown on the diagram in natural language. An example of this prompt can be found in our online appendix\footnote{The appendix is available at \url{https://zenodo.org/records/10301347}.}.
\end{enumerate}

We developed a simple application that allowed us to submit all diagrams to the model with their corresponding prompts.
We also automated the saving of the resulting classes to ease our work.

\subsection{Evaluation}

We created an evaluation strategy to evaluate all of the generated class files manually.
During this process, we assigned points for the generated source code elements based on the following criteria (we tried to assign a criterion to one exact element on the diagram):

\begin{enumerate}
    \item A class in the diagram exists in the source code: \textit{1 point}.
    \item Every data member and method of the class: \textit{1 point}.
    \item Correct visibility of any field or method: \textit{1 point}.
    \item Methods that have an implementation (at least partial): \textit{1 point}\footnote{We considered existing but empty default constructors as a valid implementation, as in several cases, developers create an empty default constructor with a purpose.}.
    In this study, any non-empty implementation (implementation that contains any source code, besides returning a placeholder value (0, false, null)) was considered partial implementation\footnote{To ensure objective evaluation, unit tests will be utilized in the future.}.
    \item Modifiers of the elements depicted in the diagram (e.g., "abstract"): \textit{1 point}.
    \item Relationships between classes (e.g., inheritance, implementation): \textit{1 point} per relationship. Since aggregations are often depicted as data members on diagrams, these will be counted only once.
\end{enumerate}

The total score represents the sum of points awarded based on these criteria.
We also measured the size of the returned source code and the associated scores for each piece of code.

\section{Preliminary Results}
\label{sec:results}

{\centering
\small
\begin{table*}[]
\caption{Scores achieved by the generated source codes}
\begin{tabular}{@{}ll|r|rrr|rrr@{}}
\toprule
\multirow{2}{*}{Type} &
  \multirow{2}{*}{Name} &
  \multirow{2}{*}{\begin{tabular}[c]{@{}l@{}}Maximum\\ score\end{tabular}} &
  \multicolumn{3}{c|}{Achieved score} &
  \multicolumn{3}{c}{Measured source lines of code} \\
                               &                     &     & Prompt 1 & Prompt 2 & Prompt 3 & Prompt 1 & Prompt 2 & Prompt 3 \\ \midrule
\multirow{10}{*}{Single-class} & Account             & 7   & 7        & 7        & 7        & 26       & 40       & 36       \\
                               & Book                & 6   & 6        & 6        & 6        & 35       & 35       & 35       \\
                               & Course              & 6   & 6        & 6        & 6        & 31       & 31       & 31       \\
                               & Employee            & 6   & 6        & 6        & 6        & 31       & 35       & 39       \\
                               & Event               & 6   & 6        & 6        & 6        & 32       & 40       & 36       \\
                               & Person              & 8   & 8        & 8        & 8        & 32       & 32       & 40       \\
                               & Product             & 8   & 8        & 7        & 8        & 41       & 33       & 34       \\
                               & Rectangle           & 7   & 7        & 7        & 7        & 40       & 31       & 28       \\
                               & Timer               & 8   & 6        & 8        & 6        & 16       & 22       & 19       \\
                               & Vehicle             & 8   & 8        & 8        & 8        & 34       & 34       & 34       \\
\midrule
\multirow{8}{*}{Multi-class}   & person\_coach\_user & 41  & 28       & 30       & 28       & 29       & 46       & 82       \\
                               & abcdef\_classes     & 123 & 68       & 35       & 108      & 44       & 50       & 44       \\
                               & book                & 155 & 80       & 86       & 124      & 106      & 113      & 161      \\
                               & knight              & 109 & 84       & 93       & 83       & 110      & 142      & 73       \\
                               & person\_habit       & 101 & 70       & 78       & 96       & 85       & 95       & 135      \\
                               & series              & 98  & 79       & 89       & 87       & 92       & 108      & 119      \\
                               & space\_station      & 101 & 85       & 83       & 81       & 102      & 104      & 86       \\
                               & witch               & 92  & 78       & 85       & 88       & 88       & 134      & 151      \\ \bottomrule
\end{tabular}
\label{tbl:res}
\end{table*}
}

The results we achieved are shown in Table~\ref{tbl:res}.
We calculated the highest possible score, presented in the ``Maximum score'' column.
This is followed by the score achieved for the prompts: simplest prompt, more detailed prompt, and finally, the most detailed prompt.
These are followed by the lengths of the source codes returned by each prompt (we considered the actual lines of code only since the model often includes comments with the source code, but not in every case. For consistency, we excluded comments from this metric).

\begin{figure}[!h]
\centering
\includegraphics[width=0.3\linewidth]{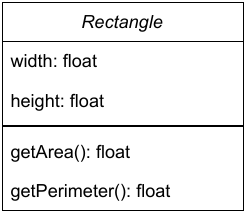}
\caption{The input diagram for example "Rectangle"}
\label{fig:rect}
\end{figure}

For example, for the "Rectangle" input, the highest possible score was 7, as the corresponding diagram showed one class (1 point), 2 fields (without visibility, 1 point each), and 2 methods (without any provided modifier, 2 point each (one if the method exists in the source code, and one if it is implemented).
Each prompt we used yielded a perfect score in this case.
The provided image input we used is shown in Figure~\ref{fig:rect}.

\vspace{-10pt}

As can be seen, for almost every single-class input, the GPT-4-Vision provided a "perfect" source class.
We note, however, that we considered every non-empty method body as a "partial implementation", hence we scored them 1 point.

In the cases of multi-class inputs, the model was not able to achieve such high scores.
For example, in the case of book\footnote{The diagrams and the generated source codes are available in our online appendix.}, the highest possible score was 155.
Our first input achieved a score of 80 (51.61\%), the second, more detailed prompt achieved a score of 86 (55.48\%), while the most detailed prompt achieved a score of 124 (80\%).
We can see that the most detailed prompt yielded a class with 161 source code lines.

In the worst-case scenario (\texttt{abcdef\_classes}, prompt 2), GPT achieved a mere 28.45\% success rate in source code generation. In contrast, in the best-case scenario (excluding the simple cases which yielded 100\%), successful source code generation reached a 95.65\% success rate (\texttt{witch}, prompt 3).

We would like to emphasize that this is the result of a manual evaluation, in which we accepted any non-empty and non-skeleton implementation as a valid implementation to avoid making subjective decisions on when a source code can be considered correct.
The production of unit tests with adequate code coverage for an objective evaluation is ongoing.

In some instances, the scores achieved by the second prompt are higher than those obtained with the third prompt.
However, in these cases, the source code generated by the third prompt is typically more extensive, and the resulting methods are more detailed (according to our scoring system, both a simple implementation and a more detailed one receive the same score).
Nonetheless, there are occasions where the model autonomously omits methods that are present in the diagram.
This issue can likely be resolved by fine-tuning the textual instructions in the prompt.
We are currently working on this issue, too.

\noindent\fbox{%
    \parbox{\linewidth}{%
\textbf{Answer to RQ}:
The GPT-4-Vision model was able to generate (partially) correct source code for an average of 88.25\% of the elements presented in the class diagram. There are simpler cases in which the model can produce perfect source code, while in more complex cases, the model can become confused.
    }%
}

\subsection{Discussion of the results}

In some instances, we observed that the source code returned by the GPT did not include all the information that was present in the diagram (for example, in the case of abcdef\_classes, prompt 2); these were indicated in comments, for example: "\textit{Other classes like BClass, CClass, DClass, EClass, and FClass will follow similarly based on the provided diagram.}".
We believe this issue could be resolved by fine-tuning our prompts, which we will definitely do in the future.

However, there were cases where the generated code contained additional functionality, such as getter/setter functions that were not present in the input diagram.
These methods were not included in the scoring, but in real-world scenarios, such methods could be beneficial in certain circumstances.
In the case of "Person", we mentioned in the third prompt that we work with email addresses, which follow a traditional format.
In the setter method for this variable, GPT generated a minimal validation, shown in Listing~\ref{lst:email}.

\begin{lstlisting}[caption={E-mail validation generated by GPT-4-Vision}, label=lst:email, language=Java]
public void setEmail(String email) {
    if (email.contains("@")) {
        this.email = email;
    } else {
        System.out.println("Invalid email format.");
    }
}
\end{lstlisting}

We observed that the visibility modifiers indicated on diagrams (denoted by "+", "-", "\#" symbols) are rarely considered by the model.
In most cases, it consistently generates private data members within a class while the methods are public. This convention could be useful when the diagram lacks specific information regarding visibility.

Regarding class diagrams containing several classes, we have also observed that the GPT can be confused.
In one such instance, the GPT generated an interface source code for a traditional class (its data members were transformed to abstract methods in this case).
Furthermore, the GPT introduced a non-existent relationship with this interface class.
The generated source code is available in our online appendix.

An interesting case can be found in the "witch" example.
The class ``FairyWitch'' inherits from an existing class and implements an interface simultaneously.
The model comprehended this and was also able to handle this case correctly, resulting in an appropriate source code.
Moreover, it used the ``@Override'' annotation for the implemented method.

We have also observed in many cases that the model attempts to create up-to-date and modern source code, e.g., utilizing streams on multiple occasions to compute minimum, maximum, and average values from a list.

\section{Future Work}
\label{sec:future}

In our plans, we aim to work with a much more diverse set of UML class diagrams, for which we have already gathered a dataset.
We would like to include a variety of UML diagrams in our study, such as scanned diagrams, hand-drawn diagrams, as well as incomplete or even damaged diagrams.
We are also working on improving the prompting we use to gain more information about how prompting affects the quality of the generated code.
Currently, we are in the process of creating unit tests for the presented examples.
Through these tests, we would like to objectively evaluate the generated source code in addition to manual validation.
We also plan to examine the code synthesis capabilities of GPT for different programming languages, including but not limited to Python and C++. Furthermore, we wish to compare the results.
And, of course, we would like to exploit a bigger aspect of UML: we would like to include use-case and sequential diagrams so that the model could better understand the intended use cases.
In the future, our plans include identifying design patterns and anti-patterns based on UML diagrams with the assistance of GPT-4-Vision.

\section{Conclusion}
\label{sec:conclusion}

In this paper, we presented our preliminary results on the capability of the latest GPT model, GPT-4-Vision, for generating source code from UML class diagrams.
We developed a framework that can send images to the model in batches using various prompts.
We demonstrated the capabilities of GPT-4-Vision using 18 UML class diagrams of varying complexity.
Based on our results, we can say that the GPT-4 has additional potential in the domain of automatically transforming UML diagrams into code.
Additionally, we observed that the detail of the given textual prompt can influence the model's performance.

\begin{acks}
The research was supported by the Ministry of Innovation and Technology NRDI Office within the framework of the Artificial Intelligence National Laboratory Program (RRF-2.3.1-21-2022-00004).
Project no. TKP2021-NVA-09 has been implemented with the support provided by the Ministry of Innovation and Technology of Hungary from the National Research, Development and Innovation Fund, financed under the TKP2021-NVA funding scheme.
\end{acks}

\bibliographystyle{ACM-Reference-Format}
\bibliography{msr.bib}


\end{document}